\documentclass[noshowpacs,amsmath,twocolumn,superscriptaddress,10pt,aps,prb]{revtex4-1} 
\usepackage{setspace}
\usepackage{amsmath}
\usepackage{breqn}
\usepackage{graphicx}
\usepackage[nearskip,margin = 0pt,caption=false]{subfig}

\usepackage{verbatim}
\usepackage{amsfonts}
\usepackage{amssymb}
\usepackage{epstopdf}
\usepackage{lipsum}
\usepackage{siunitx}
\usepackage{xcolor}
\usepackage{array}
\usepackage[normalem]{ulem}
\usepackage{layouts}
\usepackage[resetlabels,labeled]{multibib}
\DeclareGraphicsExtensions{.pdf,.eps,.png,.jpg,.mps}
\usepackage{etoolbox}

\begin{document}
\title{Inverse-Designed Silicon Nitride Nanophotonics}
\author{Toby Bi$^{1,2,\dagger}$, Shuangyou Zhang$^{1,3,\dagger}$, Egemen Bostan$^{4}$, Danxian Liu$^{4}$, Aditya Paul$^{4,5}$, Olga Ohletz$^{1}$, Irina Harder$^{1}$,\\ Yaojing Zhang$^{1,6}$, Alekhya Ghosh$^{1,2}$, Abdullah Alabbadi$^{1,2}$, Masoud Kheyri$^{1,2}$, Tianyi Zeng$^{4}$, Jesse Lu$^{7}$\\ Kiyoul Yang$^{4,*}$, and Pascal Del'Haye$^{1,2,*}$\\
\vspace{+0.05 in}
$^1$Max Planck Institute for the Science of Light, Staudtstraße 2, Erlangen, Germany.\\
$^2$Department of Physics, Friedrich-Alexander Universit\"at Erlangen-N\"urnberg, Staudtstraße 7, Erlangen, Germany.\\
$^3$Department of Electrical and Photonics Engineering, Technical University of Denmark, Kgs., Lyngby, Denmark.\\
$^4$John A. Paulson School of Engineering and Applied Sciences, Harvard University, Cambridge, MA, USA.\\
$^5$Department of Electrical Engineering and Computer Science, Massachusetts Institute of Technology, Cambridge, MA, USA.\\
$^6$School of Science and Engineering, The Chinese University of Hong Kong, Shenzhen, Guangdong, China.\\
$^7$SPINS Photonics Inc, Hollister, CA, USA.\\
$^\dagger$: These authors contributed equally to this work.\\
$^*$Corresponding authors: pascal.delhaye@mpl.mpg.de, kiyoul@seas.harvard.edu\\
}

\begin{abstract}
\noindent 
\color{black} 

\textbf{Silicon nitride photonics has enabled integration of a variety of components for applications in linear and nonlinear optics, including telecommunications, optical clocks, astrocombs, bio-sensing, and LiDAR. With the advent of inverse design -- where desired device performance is specified and closely achieved through iterative, gradient-based optimization -- and the increasing availability of silicon nitride photonics via foundries, it is now feasible to expand the photonic design library beyond the limits of traditional approaches and unlock new functionalities. In this work, we present inverse-designed photonics on a silicon nitride platform and demonstrate both the design capabilities and experimental verification by realising precisely tailored wavelength-division multiplexers, mode-division multiplexers, and high-\textit{Q} resonators with controllable wavelength range and dispersion. This demonstrates inverse-designed enhanced manipulation of orthogonal bases of light. Furthermore, we use these inverse-designed structures to form optical cavities that hold promise for on-chip nonlinear and quantum optics experiments.}
\end{abstract}

\maketitle

 \begin{figure*}[t!]
    \centering
    \includegraphics[width=0.9\linewidth]{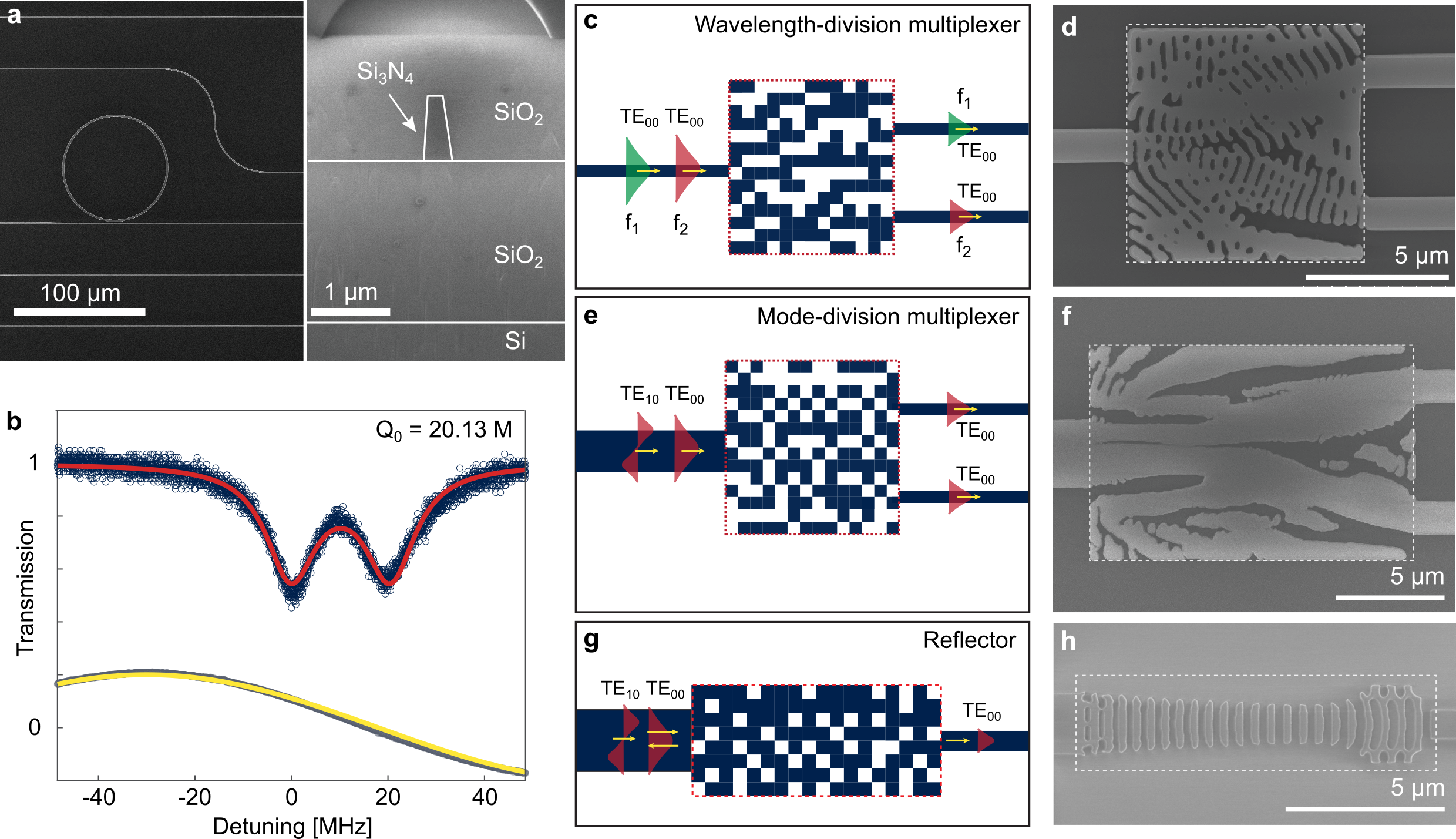}
    \caption{\textbf{Optimised silicon nitride nanophotonics.} (\textbf{a}) Scanning Electron Micrograph (SEM) images of silicon nitride photonics (top view of microring resonators (left) and cross-sectional image of waveguide (right)). (\textbf{b}) High-resolution zoom-in scan of the fundamental TE modes, with intrinsic quality factor (Q$_{0}$) indicated. M = million. The red curve is a Lorentzian fit. The yellow sinusoidal signal is a frequency calibration scan using a fibre Mach–Zehnder interferometer. (\textbf{c-d}) Generalised functionality of a 3-port wavelength-division multiplexer device and a top view SEM image of the device. Same as \textbf{c-d} but for (\textbf{e-f}) mode-division multiplexer and(\textbf{g-h}) multi-mode reflector. Dashed boxes indicate the optimisation area in both the schematics (left) and the SEM images (right).}
    \label{fig:fig1}
\end{figure*}


\section{Introduction}

\color{black} 

The development of complementary metal-oxide semiconductor processes and materials for integrated circuits has been a critical backbone in the emergence of integrated photonics. Among the most mature materials of the standard photonics platform \cite{susilicon2020}, silicon nitride has the excellent balance of large transparency window, relatively high Kerr nonlinearity, low linear and nonlinear losses (free-carrier and two-photon absorption free), high mode confinement and a low thermo-optic coefficient. Leveraging these advantages, silicon nitride-based photonic integrated circuits have raced ahead in demonstrating telecommunications \cite{marin-palomomicroresonatorbased2017, fulophighorder2018,Bergman:2023:NaturePhotonics,yangmultidimensional2022}, optical clocks\cite{ACES:2019:Optica}, LiDAR \cite{trochaultrafast2018,Kippenberg:2020:Nature}, astrocombs \cite{obrzudmicrophotonic2018, suhsearching2019}, amplifiers \cite{liuphotonic2022, gaafarfemtosecond2024}, and spectroscopy \cite{duttonchip2018, yugasphase2018} on an integrated platform.

In conjunction with the development of integrated photonics, new approaches to photonic device design have been introduced, including photonic inverse design. The inverse design, gradient-based optimisation method \cite{jensentopology2011,moleskyinverse2018}, has been demonstrated for wavelength-selective \cite{piggottinverse2015,suinverse2018,piggottinversedesigned2020,xuinversedesigned2021}, mode-selective (spatial and polarisation) \cite{shenintegratednanophotonics2015,frellsentopology2016,changultracompact2018,piggottinversedesigned2020,yangmultidimensional2022}, and resonant devices \cite{yugenetically2017,Vuckovic:2020:NaturePhotonics,ahnphotonic2022,yanginversedesigned2023}. Together, these components have demonstrated applications in particle accelerators \cite{sapraonchip2020,haeuslerboosting2022}, optical communications \cite{yangmultidimensional2022}, LiDAR\cite{Vuckovic:2020:NaturePhotonics,Vuckovic:2021:ACSPhotonics} and flat-optics \cite{backercomputational2019,liinverse2022,manninverse2023}. Most inverse-designed structures have been implemented on the standard silicon photonic platform, however more recently, there have been experimental demonstrations on diamond\cite{Vuckovic:2019:NatureCommunications}, silicon carbide\cite{yanginversedesigned2023}, and lithium niobate\cite{Wang:2023:ACSPhotonics,Kwon:2025:Nanophotonics} material platforms.

In this work, inverse-designed nanophotonics is applied to a low-loss silicon nitride platform -- specifically, a foundry-compatible, high-quality, thick silicon nitride layer (400\,nm to 800\,nm). In particular, 800-nm-thick silicon nitride photonics\cite{lukeovercoming2013,Kippenberg:2021:NatureCommunications} has demonstrated exceptionally low waveguide transmission loss and strong mode confinement. However, to the best of our knowledge, experimental demonstrations of inverse-designed photonics on this platform have remained elusive. As experimental validations, we separately demonstrate wavelength- and mode-division multiplexer structures operating around the telecommunications wavelength band. Both structures exhibit insertion losses of approximately \textminus{}2\,dB at their central operational wavelengths and crosstalk levels below \textminus{}10\,dB. Due to the lower index contrast compared to silicon and other photonic materials, silicon nitride photonics requires a relatively larger design area. In this study, we benchmark device performance for both wavelength- and mode-division multiplexers as a function of optimization area. 

Furthermore, photonic inverse design is applied to optimise optical resonators that define standing waves at resonance frequencies using waveguide-integrated reflectors. The maximum loaded finesse is measured to be approximately 162, corresponding to a \textit{Q}-factor of 0.21\,million. This design framework can potentially be extended to dispersion and dissipation engineering. The inverse-designed 800-nm-thick silicon nitride photonics platform is readily applicable for integration with nonlinear and quantum photonics.

\begin{figure*}[t!]
    \centering
    \includegraphics{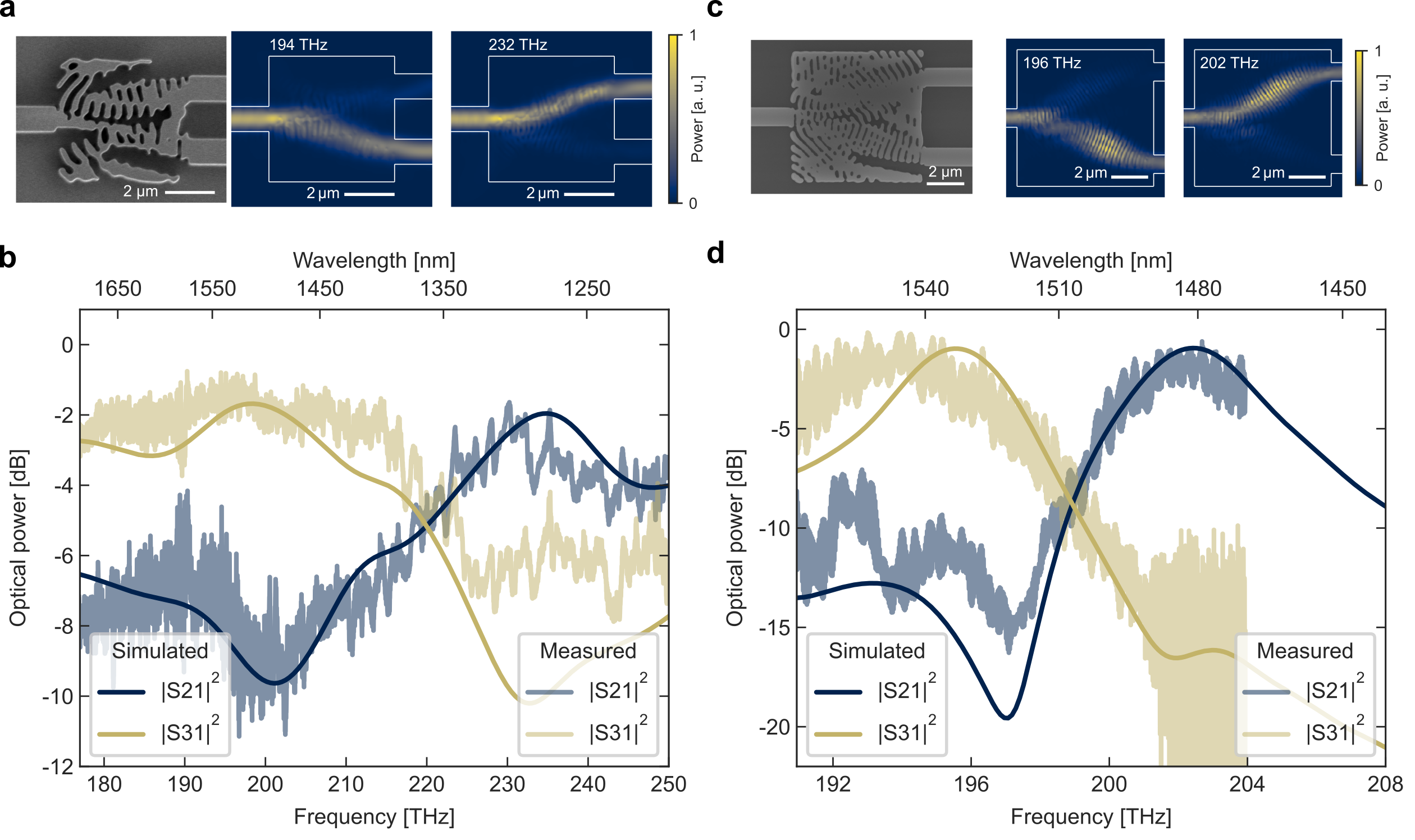}
    \caption{\textbf{Inverse-designed wavelength-division multiplexer.} (\textbf{a}) Left: SEM image of 194/232\,THz device before encapsulation with a footprint of 5\texttimes{}5\,\textmu{}m\textsuperscript{2}. Centre and right: Simulated power distribution obtained using 3D-FDTD simulations of an inverse-designed WDM with broadband input from the left waveguide measured at 194\,THz (centre) and 232\,THz (right). The input and output waveguides are 1\,\textmu{}m in width and the input mode is a quasi-TE fundamental mode. (\textbf{b}) Numerically simulated (opaque curves) and experimentally measured (transparent curves) normalised transmission of the device for fundamental TE mode input. The blue curves represent S21 while the gold curves show S31 with at best \textminus{}7\,dB of measured crosstalk in the 194\,THz channel and \textminus{}3\,dB in the 232\,THz channel. The device's 3\,dB bandwidth of the simulated \textit{S}-parameter is greater than 30\,THz for both frequency bands. (\textbf{c}) Left: SEM image of 196/202\,THz device before encapsulation with a device area of 8\texttimes{}8\,\textmu{}m\textsuperscript{2}. Centre and right: Simulated power distribution obtained using 3D-FDTD simulations of an inverse-designed WDM with broadband input from the left waveguide measured at 196\,THz (centre) and 202\,THz (right). (\textbf{d}) Numerically simulated (opaque curves) and experimentally measured (transparent curves) normalised transmission of the 196/202\,THz device. The blue curves represent the 202\,THz channel while the gold curve shows the 196\,THz channel with a peak \textminus{}11\,dB measured crosstalk for both channels. The insertion loss in both channels is around \textminus{}1.5\,dB.}
    \label{fig:fig2}
\end{figure*}

\begin{figure*}[t!]
    \centering
    \includegraphics{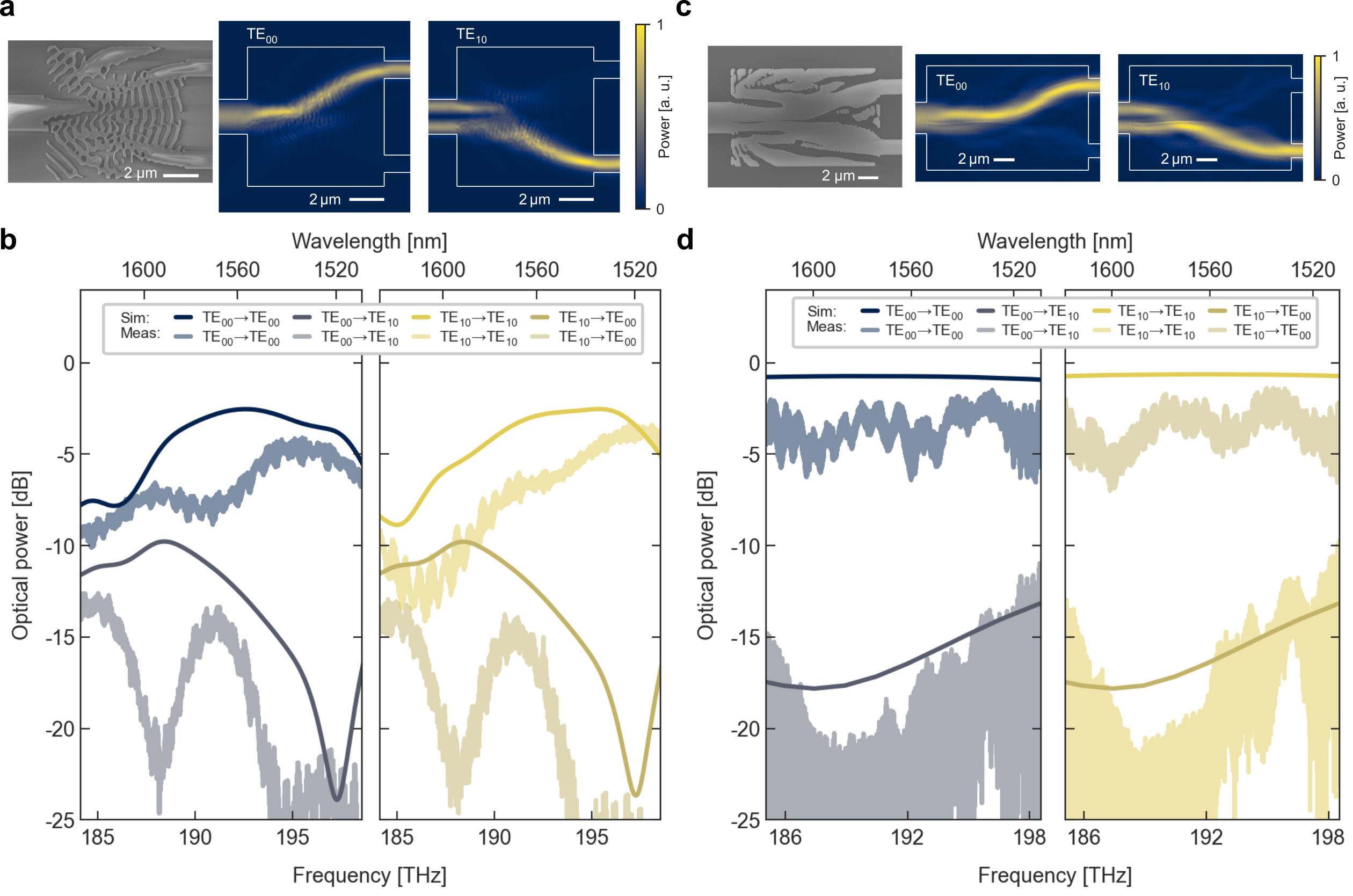}
    \caption{\textbf{Inverse-designed mode-division multiplexer (MDM).} (\textbf{a}) Left: SEM image of the 8\texttimes{}8\,\textmu{}m\textsuperscript{2} MDM. Charging effects are observed in the SEM image on the multi-mode waveguide. Centre and left: Simulated power distribution obtained using 3D-FDTD simulations of an inverse-designed MDM. The input multi-mode 2-\textmu{}m-wide waveguide is excited with the TE\textsubscript{00} (TE\textsubscript{10}) mode and is routed to the top (bottom) 1-\textmu{}m-wide output waveguide. (\textbf{b}) Response of two back-to-back inverse-designed MDMs where the opaque and transparent curves represent the simulated and measured devices, respectively. For the TE\textsubscript{00} mode (left panel), the blue and grey curves are the passing and suppressing channels, respectively, while for the TE\textsubscript{10} mode (right panel), the yellow and gold curves are the passing and suppressing channels, respectively. Here, a maximum of \textminus{}18\,dB crosstalk is achieved at 197\,THz while at least \textminus{}3\,dB crosstalk is achieved across the entire C-band. (\textbf{c}) Left: SEM image of the 15\texttimes{}10\,\textmu{}m\textsuperscript{2} MDM. Centre and left: Simulated power distribution where the input multi-mode 3-\textmu{}m-wide waveguide is excited with the TE\textsubscript{00} (TE\textsubscript{10}) mode and is routed to the top (bottom) 1.5-\textmu{}m-wide output waveguide. (\textbf{d}) Response of a similarly constructed back-to-back inverse-designed MDMs. Here, a measured minimum of \textminus{}17\,dB crosstalk is achieved at 189\,THz with 10\,THz bandwidth of at least \textminus{}10\,dB measured crosstalk and a noticeable reduction in crosstalk variation.}
    \label{fig:fig3}
\end{figure*}

\section{Results}

Inverse-designed devices are optimised using a gradient-based algorithm \cite{lunanophotonic2013,sunanophotonic2020}. The platform used across all device optimisation in this work is a Si\textsubscript{3}N\textsubscript{4} core and a symmetric SiO\textsubscript{2} cladding \cite{zhanglowtemperature2023,lukeovercoming2013,Kippenberg:2016:Optica,Kippenberg:2021:NatureCommunications} as shown in the right of Fig. \ref{fig:fig1}a. To set a benchmark for the propagation losses, we fabricate ring resonators with 40-\textmu{}m-radius on the same chip as the optimised structures (Fig. \ref{fig:fig1}a left). The resonators are measured with a telecommunications band, narrow linewidth, and low power scanning continuous-wave (CW) laser calibrated using a fibre Mach-Zehnder interferometer. Figure \ref{fig:fig1}b displays a single fundamental transverse-electric (TE\textsubscript{00}) resonance (blue open circles) co-measured with a fibre Mach-Zehnder interferometer period (black circles are measured and yellow trace is a fit). The resonance is fitted with a Lorentzian (red trace) that indicates a measured intrinsic \textit{Q}-factor (\textit{Q}\textsubscript{0}) corresponding to over 20 million, highlighting the platform's capabilities in providing both compactness and ultralow-loss.

In this work, we present three inverse-designed devices utilising the thick silicon nitride material platform; a wavelength-division multiplexer (WDM), mode-division multiplexer (MDM), and a reflector. Each device functionality is schematically shown in Fig. \ref{fig:fig1}c, e, and g for the WDM, MDM, and reflector, respectively. For an optimised 3-port WDM (Fig. \ref{fig:fig1}c, d), a broadband fundamental transverse-electric (TE\textsubscript{00}) input at the left input is separated into two spectrally separated channels centred at f\textsubscript{1} and f\textsubscript{2} while preserving the spatial mode. Next, a separate device cleanly separates the TE\textsubscript{00} and TE\textsubscript{10} spatial modes launched from the common waveguide to TE\textsubscript{00} modes in two single-mode waveguides in Fig. \ref{fig:fig1}e, f. Finally, to demonstrate the potential of the platform for nonlinear and quantum optics in a compact platform, microresonators formed using multi-mode reflectors are designed and fabricated. The device simultaneously supports high reflectivity for only the TE\textsubscript{00} mode at the multi-mode waveguide interface at the right hand side of Fig. \ref{fig:fig1}g, h and also suppresses TE\textsubscript{10} output through the reflector. Figure \ref{fig:fig1}d, f, and h displays top-view, SEM images of the complex device topology transferred into the thick silicon nitride material platform.

\vspace{1ex}

\subsection{Wavelength-division multiplexers}

For the WDMs, two 3-port devices -- one operating at  194/232\,THz and the other at 196/202\,THz -- are presented in this section. A 5\texttimes{}5\,\textmu{}m\textsuperscript{2} design area is optimised for the 194/232\,THz frequency channels. The fundamental TE mode is coupled into the WDM structure and is de-multiplexed into two separate 1\,\textmu{}m-wide output waveguides, both supporting the fundamental TE mode. The input waveguide is also 1\,\textmu{}m wide. To benchmark optimisation performance across different design areas, another WDM device is optimised with a larger footprint (8\texttimes{}8\,\textmu{}m\textsuperscript{2}), and the frequency channels are positioned closer together (196/ 202\,THz). Additionally, the operational bands of the 194/232\,THz WDM overlap with those of the C- and O-bands, making it relevant for communication applications. The 196/202\,THz WDM is applicable to erbium-based silicon nitride photonics, where a 202\,THz pump can be efficiently multiplexed with a 196\,THz signal for amplification and lasing\cite{liuphotonic2022, gaafarfemtosecond2024}.

3D finite-difference time-domain (FDTD) simulations show the device performance at the spectral centre of the passbands of the output right waveguides after excitation of the fundamental quasi-TE mode of input left waveguide. The centre and right panels of Fig. \ref{fig:fig2}a and \ref{fig:fig2}c shows light propagation at the two target frequencies effectively taking two unique paths without significant scattering from the sub-wavelength features. The simulated channel transmission and crosstalk of the 194/232\,THz WDM and the 196/202\,THz WDM correspond to \textminus{}2\,dB (194\,THz) / \textminus{}2\,dB (232\,THz) / \textminus{}1\,dB (196\,THz) / \textminus{}1\,dB (202\,THz) and \textminus{}8\,dB (194\,THz) / \textminus{}8\,dB (232\,THz) / \textminus{}19\,dB (196\,THz) / \textminus{}17\,dB (202\,THz), respectively. In the simulation, fabrication imperfections are not taken into account. 

For experimental characterisations, a commercial frequency comb fibre laser is coupled onto the chip via a lensed fibre. The lensed fibre is aligned to excite the TE\textsubscript{00} mode. Alignment is performed with a CW laser before switching to the frequency comb source. Light is out-coupled via another lensed fibre and monitored on an optical spectrum analyser. To account for spectrally dependent insertion loss and material-induced loss, the spectral density through the WDM is normalised against a straight waveguide absent of any devices nearby. SEM images of the final fabricated device before final SiO\textsubscript{2} encapsulation are also shown in the left panels of Fig. \ref{fig:fig2}a and Fig. \ref{fig:fig2}c for the 194/232\,THz and 196/202\,THz device, respectively.

Figure \ref{fig:fig2}b shows the numerically simulated (opaque curves) and experimentally measured (transparent curves) scattering parameters (\textit{S}-parameters) of the device by pumping through input port 1 and normalised to the transmission of a straight waveguide nearby. The yellow and blue curves represent the 194\,THz and 232\,THz channels, respectively. The predicted 3-dB bandwidth surpasses 30\,THz for both the 194\,THz and 232\,THz band with a largest measured channel crosstalk of around \textminus{}8\,dB in the C-band and \textminus{}8\,dB in the O-band. Experimental characterisation of the device yields \textminus{}7\,dB lensed fibre to lensed fibre insertion loss with around \textminus{}1.9\,dB loss through the WDM. The performance of the fabricated device relative to the numerically optimised design has degraded where the crosstalk has increased to around \textminus{}7\,dB in the C-band and around \textminus{}3\,dB in the O-band but the overall channel separation still remains. This can be attributed to fabrication defects like imperfect sidewall angles and incomplete gap fillings.

The numerically simulated and experimentally measured performance of the 196/202\,THz WDM is displayed in Fig. \ref{fig:fig2}d with opaque and transparent curves, respectively. For both channels, the simulated performance exceeds \textminus{}5\,dB crosstalk for over 5\,THz for the 196\,THz channel and over 10\,THz for the 202\,THz channel. The maximum simulated crosstalk is \textminus{}13\,dB for both channels. The experimentally measured crosstalk closely matches the simulated performance with only slight degradation and an insertion loss of \textminus{}1.5\,dB.

\begin{figure*}[t!]
    \centering
   \includegraphics{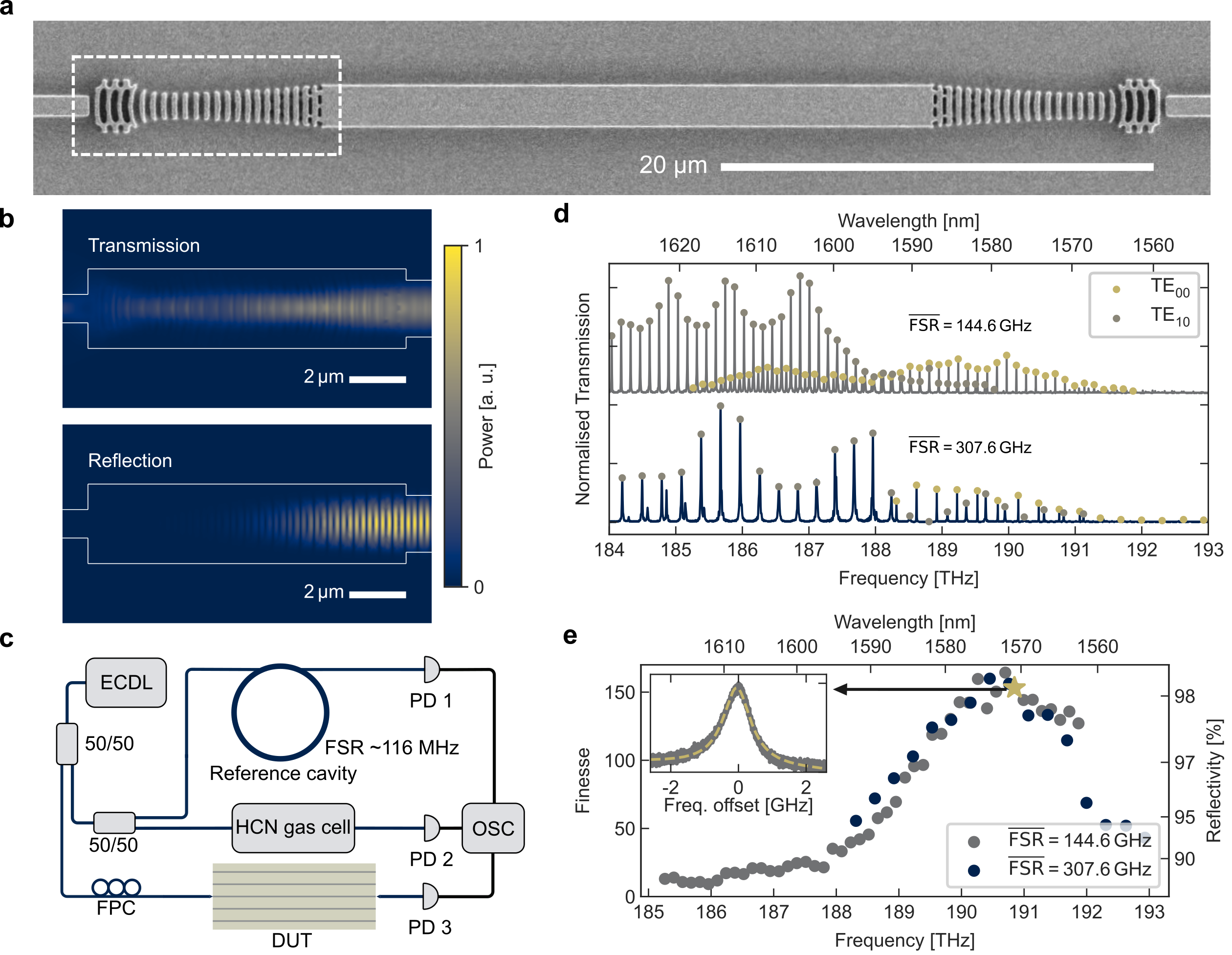}
    \caption{\textbf{Inverse-designed Fabry-P\'erot cavities.} (\textbf{a}) SEM image of a Fabry-P\'erot microresonator formed with two inverse-designed reflectors sandwiching a 2\,\textmu{}m-wide straight waveguide section with an average FSR around 307.6\,GHz. (\textbf{b}) Top and bottom panel display the power map at frequencies where light from the right-hand waveguide is transmitted through and reflected from the reflector.
    (\textbf{c}) Schematic of the linear spectroscopy setup for Fabry-P\'erot cavities. The reference cavity and the HCN gas cell are used for relative and absolute calibration, respectively, and are measured simultaneously with the DUT during every characterisation run. ECDL: external cavity diode laser, FPC: fibre polarisation controller, HCN: hydrogen cyanide, DUT: device under test, PD: photodetector, and OSC: oscilloscope. (\textbf{d}) Resonance spectrum of the TE\textsubscript{00} (gold) and TE\textsubscript{10} (grey) mode families of Fabry-P\'erot cavities with two FSRs, 144.6\,GHz (upper) and 307.6\,GHz (lower). The cavities are measured in transmission, coupling light through both reflectors. (\textbf{e}) Finesse of the FP cavities and average reflectivity of a single mirror for both cavities. A maximum finesse of 162 is measured corresponding to a reflectivity of 98.5\,\%. The inset displays a resonance at 190.8\,THz in the 144.6-GHz-FSR resonator in transmission with a loaded linewidth of 950\,MHz.}
    \label{fig:fig4}
\end{figure*}

\subsection{Mode-division multiplexers}

In this section, we present the experimental demonstration of a mode-division multiplexer. Spatial modes -- another set of orthogonal channels -- not only enable multi-dimensional data transmission\cite{yangmultidimensional2022} but also unlock new functionalities for efficient optical modulation\cite{Lipson:2020:Optica} and nonlinear multi-mode photonics\cite{Wright:2022:NaturePhysics}. To benchmark optimisation performance in this study, we compared device performance not only across different optimization areas but also with varying layer thicknesses.

An initial unoptimised area of 8\texttimes{}8\,\textmu{}m\textsuperscript{2} with a single 2-\textmu{}m-wide waveguide on one side and two 1-\textmu{}m-wide waveguides on the opposite side. Excitation from the 2\,\textmu{}m waveguide at 193\,THz is either the TE\textsubscript{00} or TE\textsubscript{10} mode family. The optimisation goal is to increase the mode overlap between the excitation mode and the TE\textsubscript{10} mode at the 1-\textmu{}m-wide output waveguides. After around 200 iterations the optimised MDM achieves crosstalk below \textminus{}21\,dB and insertion loss of \textminus{}4\,dB for a thickness of 400\,nm as shown in the opaque curves in Fig. \ref{fig:fig3}b. Increasing the footprint to 15\texttimes{}10\,\textmu{}m\textsuperscript{2} and layer thickness to 800\,nm not only reduces the insertion loss to sub \textminus{}1\,dB but increases the operational bandwidth while maintaining a crosstalk of \textminus{}17\,dB.

The final optimised device is fabricated and tested. The MDM device is designed to route the TE\textsubscript{00} and the TE\textsubscript{10} modes from a 2-\textmu{}m-wide common multi-mode waveguide into two waveguides of 1\,\textmu{}m width with peak efficiency at around 198\,THz as shown in full 3D FDTD simulations in the centre and left panel of Fig. \ref{fig:fig3}a. To evaluate the device performance, the MDM is measured in a back-to-back configuration where two MDMs are connected via a multi-mode common 100-\textmu{}m-long waveguide (500-\textmu{}m-long for larger MDM). This ensures the excitation of the desired spatial mode family in the second MDM. The larger footprint, 15\texttimes{}10\,\textmu{}m\textsuperscript{2}, MDM has a nominal multi-mode, 3\,\textmu{}m-wide common waveguide and two 1.5\,\textmu{}m-wide single-mode waveguides. The intensity maps for the larger MDM are shown in the centre and right panels of Fig. \ref{fig:fig3}c. SEM images of the fabricated device are shown in the left panels of Fig. \ref{fig:fig3}a and \ref{fig:fig3}c.

MDM characterisation is performed in the C- and L-band. To account for spectrally dependent fibre to chip insertion losses and waveguide propagation losses, the measured spectral response of the MDM is normalised to a nearby straight waveguide. The average insertion loss of a single MDM device is around \textminus{}2\,dB for both MDM sizes. Figure \ref{fig:fig3}b and Fig. \ref{fig:fig3}d displays the experimentally measured response (transparent curves) from a back-to-back configuration overlaid with the numerically simulated response (opaque curves) for the 8\texttimes{}8\,\textmu{}m\textsuperscript{2} and 15\texttimes{}10\,\textmu{}m\textsuperscript{2} MDMs, respectively. Here, the smaller MDM achieves a mode crosstalk of \textminus{}18\,dB measured between the TE\textsubscript{00} and TE\textsubscript{10} channels at around 194.5\,THz, which matches well with simulations. Similarly, the larger, 15\texttimes{}10\,\textmu{}m\textsuperscript{2}, MDM device experimentally measured performance is as expected when compared to the numerically simulated performance. However, the crosstalk variation across the C- and L-band is only \textminus{}1\,dB compared to \textminus{}15\,dB for the 8\texttimes{}8\,\textmu{}m\textsuperscript{2} footprint MDM.

\vspace{1ex}

\subsection{Microresonators}
Fabry-P\'{e}rot (FP) resonators are constructed from inverse-designed reflectors. In recent years, FP-type resonators on the integrated platform have gained increasing attention as the reflectors present an opportunity to engineer both the dispersion and dissipation. These nanophotonic refractive index modulations have been induced using a fish-bone like structure \cite{yuphotoniccrystalreflector2019}, sinusoidal modulations \cite{wildidissipative2023}, and holes \cite{nardiintegrated2024} where threshold powers are low enough due to low scattering to observe four-wave mixing and temporal cavity solitons. Traditionally-designed reflectors are applied to the silicon nitride \cite{yuphotoniccrystalreflector2019,xieonchip2020,wildidissipative2023,zhangfabryperot2023} and gallium phosphide \cite{nardiintegrated2024} material platforms. The capability of inverse-design to optimise reflectors for multiple parameters has been demonstrated in silicon \cite{yugenetically2017,ahnphotonic2022}, and silicon carbide \cite{yanginversedesigned2023}. Compared to classically designed reflectors, inverse-designed reflectors combine the function of higher-order mode filtering, mode tapering, and high reflections onto a compact footprint \cite{ahnphotonic2022}.

Inverse-designed reflectors (see dashed box on the left-hand side of Fig. \ref{fig:fig4}a) are designed for multi-mode cavity waveguides to reduce the interaction cross section between the propagating field and the scattering sidewall. A design area of 11\texttimes{}2.8\,\textmu{}m\textsuperscript{2} with a 2\,\textmu{}m-wide intracavity waveguide and 1\,\textmu{}m wide out-of-cavity waveguide is optimised. The optimisation requires 200 iterations of finite-difference frequency-domain simulations to exceed 92\,\% reflectivity at 1475\,nm, 1525\,nm, 1575\,nm, and 1625\,nm wavelengths for the fundamental TE mode family. To confirm the complete spectral response of the inverse-designed reflector, full 3D FDTD simulations using Lumerical are performed. Figure \ref{fig:fig4}b shows the in-plane response of the reflector power at transmission frequencies (centre) and reflecting frequencies (below), respectively when excited from the intracavity (right) side. At a frequency within the passband of 160\,THz, outside of the high reflectivity region of the mirror, light passes through the reflector from the inside (right) to outside of the cavity (left) while at 191\,THz, the majority of the optical field is reflected. An FP microresonator is simply formed by placing two mirrored inverse-designed reflectors on either side of a waveguide as shown in Fig. \ref{fig:fig4}a.

Experimentally, the fabricated FP devices are characterised using tunable diode laser spectroscopy \cite{zhangonthefly2024}, schematically shown in Fig. \ref{fig:fig4}c. Further details are in the Materials and Methods section. The finesse and single mirror average reflectivity of the TE\textsubscript{00} mode family for both cavity FSRs are shown in Fig. \ref{fig:fig4}d. The maximum loaded finesse is 162 located at around 190.7\,THz for both cavities, corresponding to a \textit{Q}-factor of around 210,000. The resonance measured in transmission at 190.8\,THz (gold star scatter point) is shown in the top left inset in Fig. \ref{fig:fig4}e with a fitted loaded linewidth of around 950\,MHz. The propagation loss is extracted by characterising ring resonators fabricated on the same chip. With knowledge of the propagation loss, the other major loss channel, the mirror reflectivity, can be determined. A radius of curvature of 100\,\textmu{}m for the ring resonators is chosen to minimise bending losses to isolate the propagation loss. Here, an average intrinsic \textit{Q}-factor of 4.7 million is measured for the TE\textsubscript{00} mode family across an 8 THz bandwidth, corresponding to an average propagation loss of 8\,dB/m. The reflectors presented here have an average reflectivity of above 97\,\% across 3\,THz with a maximum reflectivity of 98.5\,\% for the TE\textsubscript{00} mode family.

\vspace{1ex}

\section{Conclusion and outlook}
In summary, we proposed, fabricated and characterised inverse-designed WDMs, MDMs, and standing-wave resonators formed with inverse-designed reflectors in thick silicon nitride platform. For the demonstrated WDM devices, we achieve channels of 194/232\,THz and 196/202\,THz in a 5\texttimes{}5\,\textmu{}m\textsuperscript{2} or 8\texttimes{}8\,\textmu{}m\textsuperscript{2} footprint. Additionally, TE\textsubscript{00} and TE\textsubscript{10} spatial modes are separated using a compact, inverse-designed MDM with a \textminus{}18\,dB crosstalk between the two separated channels at a footprint of 8\texttimes{}8\,\textmu{}m\textsuperscript{2} and 15\texttimes{}10\,\textmu{}m\textsuperscript{2}. Finally, FP cavities are formed with inverse-designed reflectors with a 11\texttimes{}2.8\,\textmu{}m\textsuperscript{2} footprint, exhibiting a Finesse exceeding 160 for an FSR of around 144.6\,GHz and 307.6\,GHz. This is equivalent to a reflectivity of above 98\,\% for a single mirror.

All inverse-designed structures demonstrated in this paper can be readily integrated into nonlinear and quantum photonic circuits. It is important to note that the thick-nitride platform is the key to achieve  most Kerr-based nonlinear and quantum photonic operations in a relatively compact surface area. Apart from the device footprint, the layer thickness requirement eases access to various dispersion conditions, and prior work using the nitride photonic platform does not meet this requirement\cite{ruizsilicon2023, ruizinversedesigned2024, ruizinversedesigned2025}. Furthermore, advancements in the design process have lifted minimum feature size constraints during foundry fabrication, increasing the accessibility of inverse-designed devices \cite{piggottinversedesigned2020, yangmultidimensional2022, schubertinverse2022}. These results demonstrate the feasibility of compact, fabrication-error-robust, customised photonic components and pave the way for scalable, high-performance integration in silicon nitride-based photonic systems.

\bibliography{Reference}

\newpage
\section{Materials and Methods}

\subsection{Inverse design}
Optimised designs are obtained using SPINS\cite{lunanophotonic2013, sunanophotonic2020}. For the larger footprint MDM (10\texttimes{}15\,\textmu{}m\textsuperscript{2} footprint) SPINS is used in conjunction with \textit{fdtd-z}. A minimum feature size of 80\,nm to 100\,nm and moderate curvature tolerance is used to accommodate for imperfections in the lithography, etching, and gap filling steps. A square mesh of 40\,nm is used to accurately simulate device performance. During each optimisation step, features are filtered to increase fabrication feasibility. After optimised devices are obtained, device performance is verified with full 3D-FDTD simulations using Lumerical. Finally, for all designs considered, any small features prone to fabrication errors are manually removed.

\subsection{Device fabrication}
For the results in Fig. \ref{fig:fig2}a-b, Fig. \ref{fig:fig3}a-b, and Fig. \ref{fig:fig4}, silicon nitride thin-film is deposited on a 525-\textmu{}m-thick silicon wafer with 3-\textmu{}m-thick thermal oxide via reactive magnetron sputtering \cite{frigglow2019, zhanglowtemperature2023}. The thickness and the refractive index of the silicon nitride thin-film can be varied by adjusting the sputtering time and the nitrogen and argon gas ratio. All devices are fabricated on films targeting stoichiometric compositions but with varying film thicknesses. The thick, sputtered silicon nitride based WDM, MDM, and reflectors are fabricated on 730\,nm, 790\,nm and 400\,nm thickness, respectively. Once the film is deposited, we spin-coat maN-2400 series negative e-beam resist, perform e-beam lithography and develop in maD-532 solution. The resist is used as an etch-mask in a reactive ion etch with an inductively coupled plasma in a CHF\textsubscript{3}/O\textsubscript{2} gas environment. Before encapsulation with SiO\textsubscript{2} cladding, the waveguides are cleaned in piranha solution. A combination of atomic layer deposition SiO\textsubscript{2} for gap filling and plasma-enhanced chemical vapour deposition  SiO\textsubscript{2} for speed is used for cladding. The chips with the Fabry-P\'erot resonators are annealed at 400\textdegree{}C to remove excess hydrogen in the cladding. Finally, the facets are exposed to the environment via manual cleaving.
Full details on the sputtered silicon nitride device fabrication can be found in \cite{zhanglowtemperature2023}. 

For the results in Fig. \ref{fig:fig1}a-b, Fig. \ref{fig:fig2}c–d and Fig. \ref{fig:fig3}c–d, the silicon nitride film is deposited on a 3-\textmu{}m-thick thermal oxide layer on a silicon wafer via low-pressure chemical vapor deposition. The target film thickness is 800\,nm for all devices shown in this work, and the lithography and etching processes follow the same procedure described in the previous paragraph. We directly deposit a SiO\textsubscript{2} top cladding via low-pressure chemical vapor deposition, and the chip was annealed at 1000\textdegree{}C.

\subsection{Microresonator Charactersiation}
Microresonators formed using inverse-designed reflectors (Fig. \ref{fig:fig4}a) are characterised using the setup in Fig. \ref{fig:fig4}c. Determining the absolute and relative frequency of the nonlinear external cavity diode laser scanning rate is achieved via a hydrogen cyanide gas cell and a calibrated, reference fibre cavity, respectively. The mode spacing of the fibre cavity in the C- and L-band is 116\,MHz and is precisely calibrated by using two radio frequency modulation calibration markers \cite{zhangonthefly2024}. The response from the gas cell, fibre cavity, and resonator devices are measured simultaneously on separate photo-detectors on a high memory depth oscilloscope. Single-mode lensed fibres are used to couple light in and out of the device-under-test. The input polarisation is controlled with a fibre polarisation controller situated before the input lensed fibre.

\vspace{1ex}

\noindent\textbf{Note}  While preparing our manuscript, we became aware of a study demonstrating inverse-designed components in a 400\,nm thick silicon nitride platform \cite{ruizultracompact2025}.

\noindent\textbf{Data availability}  The data that support the plots within this paper and other findings of this study are available from the corresponding author upon reasonable request.

\noindent\textbf{Code availability}  The code used to produce the plots within this paper is available from the corresponding author upon reasonable request.

\section*{Acknowledgments}\vspace{-0.1in}
The authors would like to acknowledge the contributions from Eduard Butzen, Florentina Gannott, Alexander Gumann, Katrin Ludwig, and Heike Schr\"oter-Hohmann from the Micro- and Nanostructuring department (MPL, Erlangen). T.B acknowledges useful discussions with Thibault Wildi (DESY, Hamburg), Alberto Nardi (IBM Research, Zurich), and Nivedita Vishnukumar (MPL, Erlangen). K.Y acknowledges discussions with Jelena Vuckovic (Stanford). This work is supported by the European Union's H2020 ERC Starting Grant No. ``CounterLight'' 756966; H2020 Marie Sklodowska-Curie COFUND ``Multiply'' 713694, Marie Curie Innovative Training Network ``Microcombs'' 812818, Max Planck Society, Max Planck School of Photonics, Max-Planck Fraunhofer Cooperation Project LAR3S, Munich Quantum Valley Project TeQSiC, and DFG Project 541267874. K.Y acknowledges support by the Under Secretary of Defense for Research and Engineering under Air Force contract number FA8721-05-C-0002. Opinions, interpretations, conclusions and recommendations are not endorsed by the US Government.\\

\noindent\textbf{Author contributions}  T.B, E.B, and A.P designed the structures. T.B, S.Z, D.L, E.B, A.P fabricated and characterised the devices with assistance from O.O, I.H, Y.Z, and T.Z. T.B, S.Z, K.Y, and P.D wrote the manuscript with inputs from all the other co-authors. P.D and K.Y supervised the project.

\end{document}